\documentclass[aps,
		pra,
		twocolumn,
		superscriptaddress,
		nofootinbib,
		10pt
		]{revtex4-1}

\usepackage{amsmath}
\usepackage{amssymb}
\usepackage{graphicx}
\usepackage{mathtools}
\usepackage{upgreek}

\makeindex
\graphicspath{
	{Figures/}
	}

\newcommand{\ybion}[1]{^{#1}$Yb${}^{+}}

\newcommand{\qed}{\nobreak \ifvmode \relax \else
      \ifdim\lastskip<1.5em \hskip-\lastskip
      \hskip1.5em plus0em minus0.5em \fi \nobreak
      \vrule height0.75em width0.5em depth0.25em\fi}

\newcommand{\ket}[1]{\left|#1\right\rangle}
\newcommand{\bra}[1]{\left\langle #1\right|}
\newcommand{\eq}{\begin{equation}}
\newcommand{\eeq}{\end{equation}}
\newcommand{\esp}{\begin{split}}
\newcommand{\spe}{\end{split}}

\newcommand{\sech}{\text{sech}}

\begin{document}

\newcommand{\var}[1]{{\operatorname{#1}}}

\newcommand{\FigureOne}{
\begin{figure}
\centering
\includegraphics[width=\columnwidth]{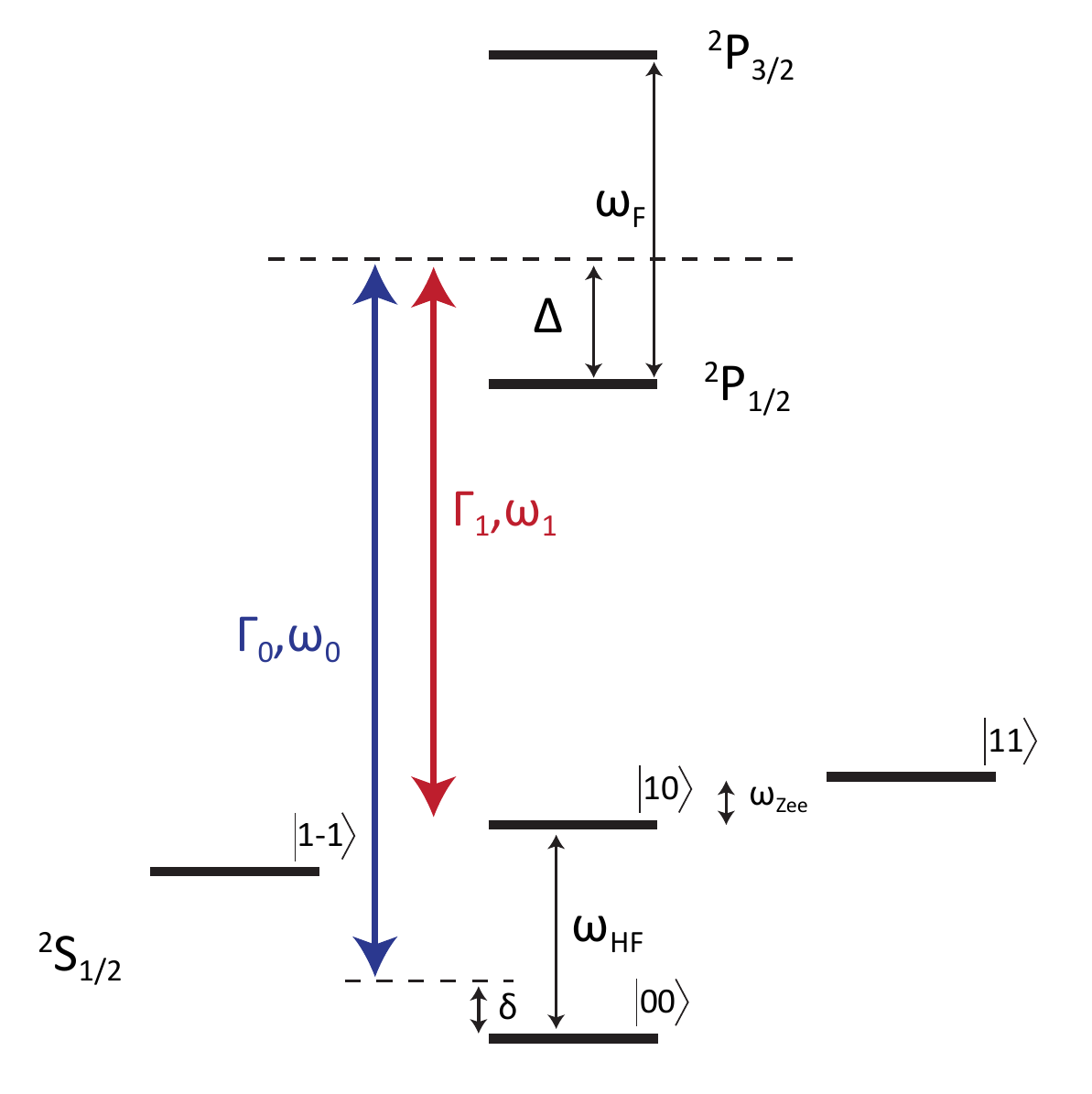}
\caption{Schematic representation of the electron energy levels of $\ybion{171}$. We encode the qubit in the ground state hyperfine clock states $\ket{00}$ and $\ket{10}$. When two phase-coherent colors of light are applied to the atom which have a beatnote approximately equal to the qubit splitting, there is an effective fourth-order differential light shift which can be much larger than the second-order differential Stark shift.}

\label{fig:AtomicLevelDiag}
\end{figure}
}

\newcommand{\FigureTwo}{
\begin{figure*}[ht]
\includegraphics[width=7in]{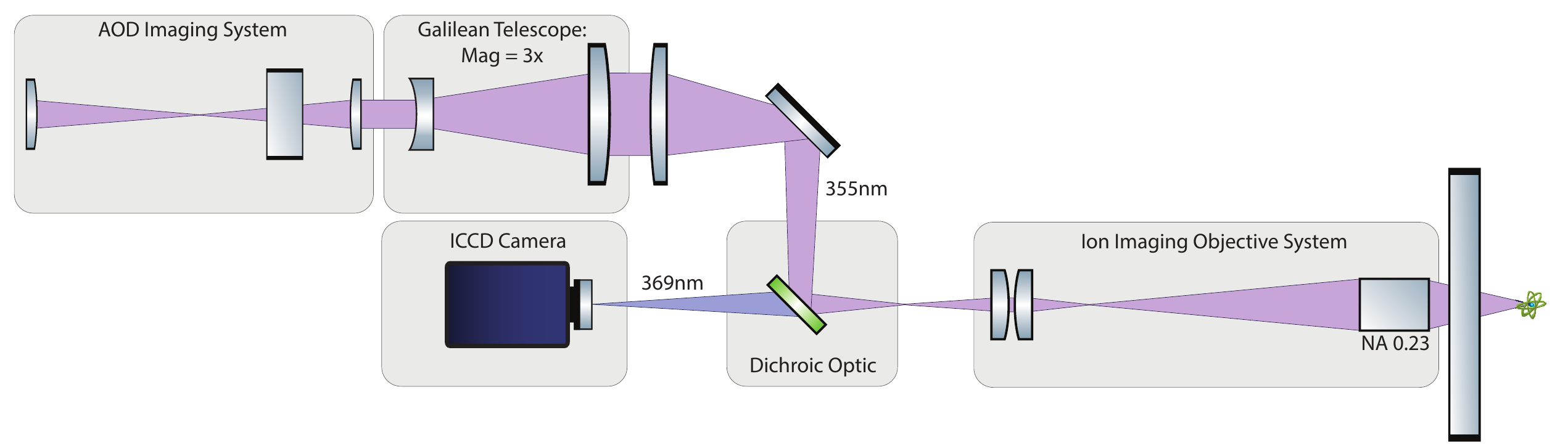}
\caption{Diagram of optics that image 355nm light onto ion chain with $<$ 3 $\upmu$m spot size, giving rise to controllable and individual-addressed Stark shifts on the qubits. This optical system utilizes a NA 0.23 objective lens for state detection of the ions at 369nm. Since the AOD is not imaged, deflections at the AOD correspond to displacement at the ions. This maps RF drive frequency to ion position, enabling control of the horizontal position of the beam.}

\label{fig:BeamPath}
\end{figure*}
}

\newcommand{\FigureThree}{
\begin{figure}
\centering
\includegraphics[width=\columnwidth]{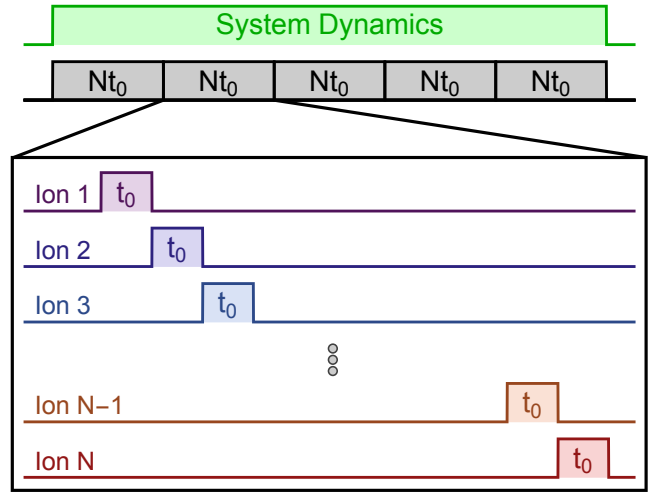}
\caption{Sketch of a typical raster pulse sequence. When the light is evenly distributed across $N$ ions, the applied fourth-order stark shift diminishes by $1/N^2$ due to the quadratic dependance on intensity. We recover a linear dependance on ion number by rastering the beam, or applying a large shift for a short time, t$_0$ sequentially to the ions. As long as each pulse chapter of length $N t_0$ is much shorter than the interaction time-scale, then the shift on each ion is then proportional to $1/N$.}
\label{fig:RasterSeq}
\end{figure}
}

\newcommand{\FigureFour}{
\begin{figure}
\centering
\includegraphics[width=.9\columnwidth]{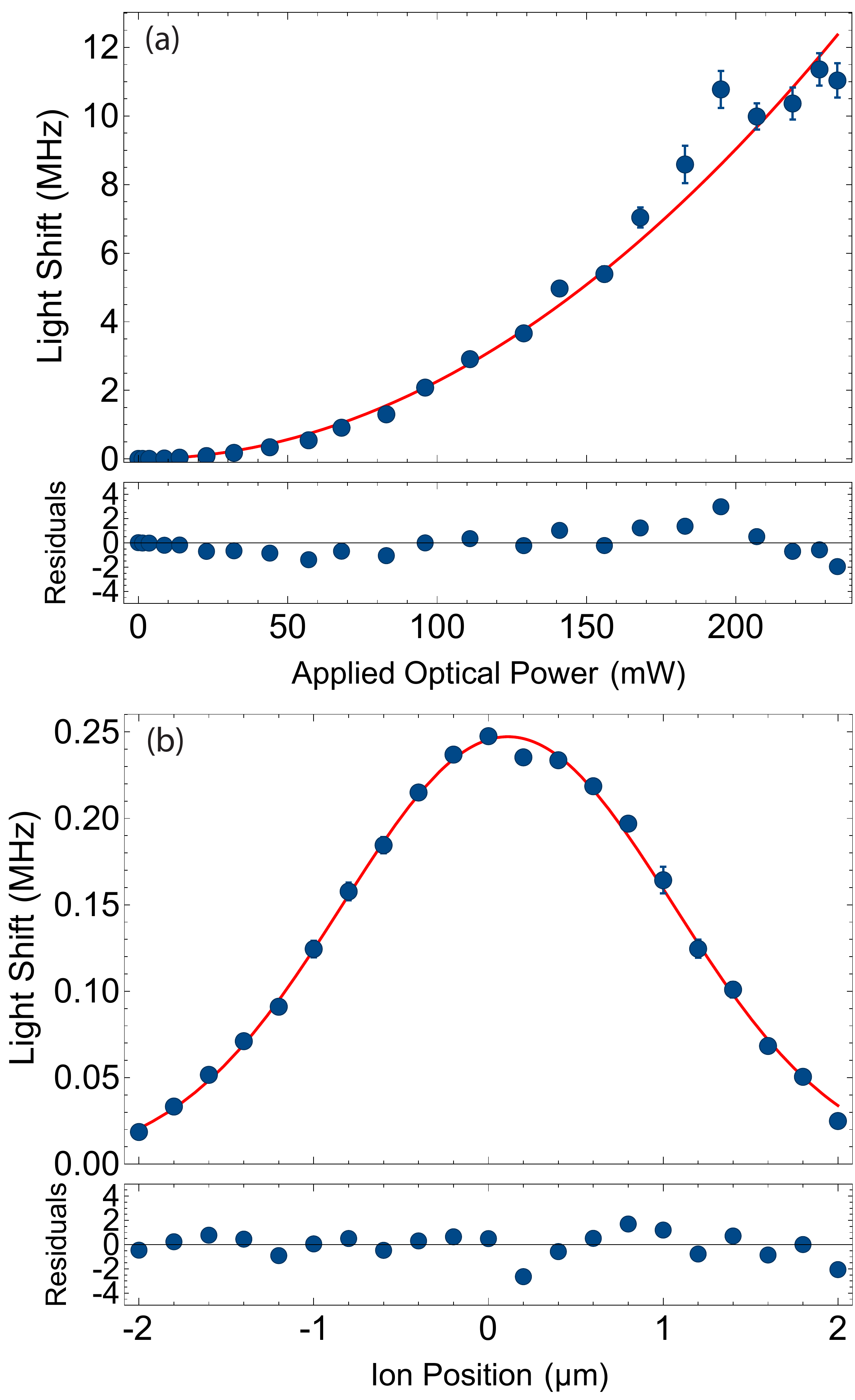}
\caption{Measured fourth-order Stark shift as a function of optical power with fit residuals (a). The clear quadratic dependance of the light shift on the applied time-averaged optical power shows that it arises from the fourth-order Stark shift. Measurement of the beam waist at the ion with fit residuals (b). By translating the ion through the beam with a fixed applied optical power of 40 mW, we extract the horizontal optical waist at the ion. We found this to be 2.68 $\upmu$m.}

\label{fig:shiftandwaist}
\end{figure}
}

\newcommand{\FigureFive}{
\begin{figure}
\centering
\includegraphics[width=\columnwidth]{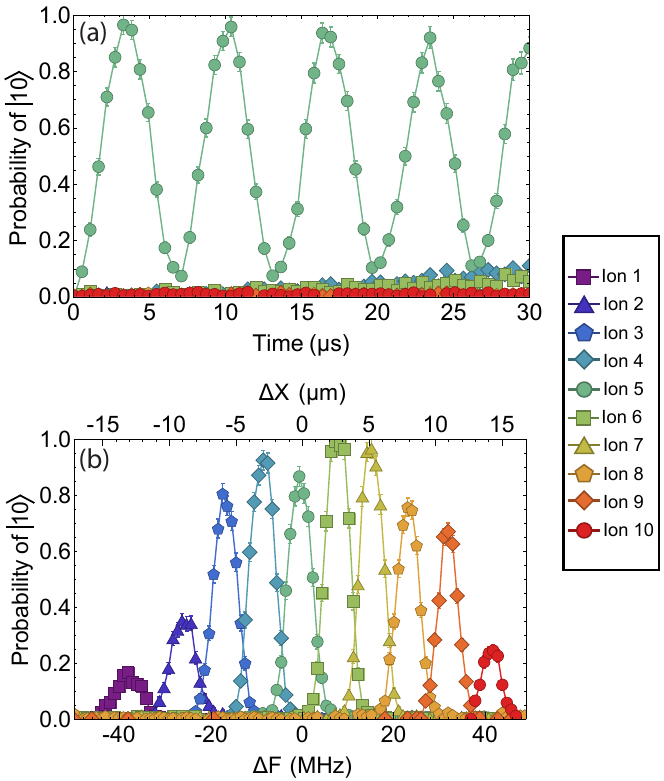}
\caption{Observed crosstalk of beam applied to one ion (a). By applying light to only ion 5 in a chain of 10, we measure the crosstalk on the nearest neighbors, ion 4 and 6, to be only 2$\%$, which is consistent with our measured horizontal beam waist and the ion separation. Individual ion signal as the beam is swept over a chain of ten ions (b). By scanning the AOD drive frequency for a fixed power and duration, we map the fourth-order Stark shift as a function of drive frequency. This corresponds to a displacement of beam position at the ion chain. The effective scanning range of the AOD is approximately 30 $\upmu$m. }

\label{fig:resolution}
\end{figure}
}

\newcommand{\FigureSix}{
\begin{figure}
\centering
\includegraphics[width=\columnwidth]{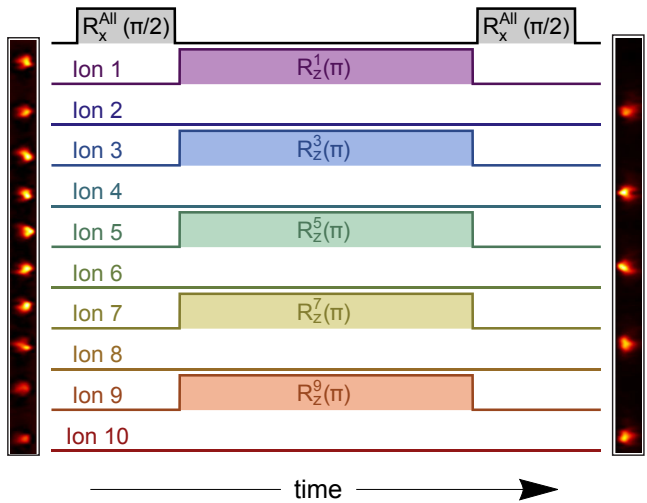}
\caption{Pulse sequence for preparing a string of 10 ions in a staggered spin configuration. All 10 ions are prepared in $\ket{00}$ and then a global $\pi/2$ pulse is applied. Depending on the state being prepared, some number of the ions have a $\pi$ phase shift applied, creating the desired configuration. A final global $\pi/2$ pulse projects the configuration back into qubit basis, completing the effective Ramsey sequence.}

\label{fig:stateprep}
\end{figure}
}

\title{Engineering Large Stark Shifts for Control of Individual Clock State Qubits}

\author{A. C. Lee}
\affiliation{Joint Quantum Institute, University of Maryland Department of Physics and National Institute of Standards and Technology, College Park, MD  20742}

\author{J. Smith}
\affiliation{Joint Quantum Institute, University of Maryland Department of Physics and National Institute of Standards and Technology, College Park, MD  20742}

\author{P. Richerme}
\affiliation{Department of Physics, Indiana University, Bloomington, IN, 47405}

\author{B. Neyenhuis}
\altaffiliation[Present address: ]{Lockheed Martin Corporation, Littleton, CO 80127, USA}
\affiliation{Joint Quantum Institute, University of Maryland Department of Physics and National Institute of Standards and Technology, College Park, MD  20742}

\author{P. W. Hess}
\affiliation{Joint Quantum Institute, University of Maryland Department of Physics and National Institute of Standards and Technology, College Park, MD  20742}

\author{J. Zhang}
\affiliation{Joint Quantum Institute, University of Maryland Department of Physics and National Institute of Standards and Technology, College Park, MD  20742}

\author{C. Monroe}
\affiliation{Joint Quantum Institute, University of Maryland Department of Physics and National Institute of Standards and Technology, College Park, MD  20742}

\date{\today}

\begin{abstract}
In quantum information science, the external control of qubits must be balanced with the extreme isolation of the qubits from the environment. Atomic qubit systems typically mitigate this balance through the use of gated laser fields that can create superpositions and entanglement between qubits. Here we propose the use of high-order optical Stark shifts from optical fields to manipulate the splitting of atomic qubits that are insensitive to other types of fields. We demonstrate a fourth-order AC Stark shift in a trapped atomic ion system that does not require extra laser power beyond that needed for other control fields. We individually address a chain of tightly-spaced trapped ions and show how these controlled shifts can produce an arbitrary product state of ten ions as well as generate site-specific magnetic field terms in a simulated spin Hamiltonian.
\end{abstract}

\maketitle

\section{Introduction}

Trapped atomic ions have emerged as one of the most promising quantum information platforms \cite{Blatt2008,Monroe2013} due to their long coherence times \cite{Bollinger1991,Fisk1997}, high fidelity readout \cite{Noek2013}, and high fidelity single \cite{Brown2011,Ballance2015,Gaebler2016} and two qubit \cite{Ballance2015,Gaebler2016} operations that are driven by external fields. Small scale quantum algorithms have even been demonstrated as the first steps toward the goal of a fault-tolerant quantum computer \cite{Monz2016, Debnath2016}. These same qualities also make atomic ions an excellent platform for quantum simulation \cite{Molmer1999,Kim2011,Blatt2012}, leveraging the long lifetimes and low noise to study dynamics that are classically intractable due to their exponential scaling with system size.

The pervasive challenge facing all quantum information platforms is the undesired interaction of the qubit with environment. In trapped ions, one such coupling to the environment is the modulation of the qubit energy splitting by stray magnetic fields. This can be circumvented by using levels whose energy difference is insensitive to magnetic fields to first order, allowing for coherence times exceeding 10 minutes \cite{Bollinger1991,Fisk1997}. Such ``clock-state" qubits are an excellent starting point for fault-tolerant quantum computation and quantum simulation \cite{NaturePhysics2012}.  For example, simulations of quantum magnetism have been performed with up to 18 spins \cite{senko2014} and with various entangling spin-spin Hamiltonians \cite{Islam2011,Lanyon2011,Islam2013,An2015,Richerme2013,Richerme2014,Jurcevic2014,Zhang2015}. However, the use of clock-state qubits by definition does not easily allow the direct generation of certain classes of Hamiltonians that are equivalent to the modulation of qubit energy splittings \cite{Lee2005}. In quantum computing, such control is also desirable for efficiently realizing universal logic gate families such as arbitrary rotations \cite{NielsenChuangBook}.

Here we propose and demonstrate the use of a fourth-order Stark shift to achieve fast, individually addressed, single-qubit rotations in a chain of $\ybion{171}$ ions.  We experimentally realize a 10 MHz shift on the qubit splitting with only moderate amounts of laser power. We exploit this control in a quantum system of 10 trapped ion clock-state qubits by preparing arbitrary initial product states and applying an independent programmable disordered splitting on each lattice site in a quantum simulation, all demonstrated with low cross-talk.

\section{Fourth-order Stark Shift Theory}

The studies reported here are performed on a linear chain of $\ybion{171}$ ions, but can be generalized to any species of clock qubits. The ions are confined using a linear radiofrequency (rf) Paul trap and the qubit is encoded in the $^2\mbox{S}_{1/2}\ket{F=0,m_f=0}$ and $^2\mbox{S}_{1/2}\ket{F=1,m_f=0}$ hyperfine clock states, denoted as $\ket{00}$ and $\ket{10}$ respectively, which have an unshifted splitting of $\omega_{HF}/2\pi=12.642821$ GHz.

We irradiate the ions using an optical frequency comb generated from a mode-locked laser with a center frequency detuned by $\Delta$ from the $^2P_{1/2}$ manifold and by $\omega_F-\Delta$ from the $^2P_{3/2}$ manifold. The laser bandwidth is much smaller than the fine structure splitting $\omega_F$ of the $P$ states and also the detuning $\Delta$. However, the laser bandwidth is much larger than the qubit splitting $\omega_{HF}$ so that the laser pulses directly drive stimulated Raman processes between the qubit states while not appreciably populating the excited $P$ states \cite{Campbell2010}. We assume that the pulse area of each laser pulse is small and has only a modest effect on the atom, and that the intensity profile for each pulse is well approximated by a hyperbolic secant envelope \cite{Campbell2010}. Under these assumptions, the $k$th comb tooth at frequency $k\nu_{rep}$ from the optical carrier has a resonant $S\rightarrow P$ Rabi frequency \cite{Mizrahi2013},
\eq
g_k=g_0\sqrt{\pi\nu_{rep}\tau}\sech(2\pi k\nu_{rep}\tau)
\eeq
where $\tau$ is laser pulse duration, $g_0^2=\gamma^2\bar{I}/2I_0$, $\bar{I}$ is the time-averaged intensity of the laser pulses, $I_0$ is the saturation intensity of the transition, and $\gamma$ is the spontaneous decay rate. Since $\sum_{k=-\infty}^\infty g_k^2 = g_0^2$, and assuming the parameters specified above, the second-order Stark shift $E^{(2)}_\alpha$ of state $\ket{\alpha}$ due to the frequency comb can be computed for an arbitrary polarization (taking $\hbar=1$) \cite{Wineland2003,Campbell2010}:
\eq\label{eq:twoStark}
\begin{split}
E^{(2)}_{00}=&\frac{g_0^2}{12}\left(\frac{1}{\Delta}-\frac{2}{\omega_F-\Delta}\right)\\
E^{(2)}_{10}=&\frac{g_0^2}{12}\left(\frac{1}{\Delta+\omega_{HF}}-\frac{2}{\omega_F-(\Delta+\omega_{HF})}\right).\\
\end{split}
\eeq
Here we neglect all excited state hyperfine splittings since they only contribute to the Stark shifts at a fractional level of $\sim 10^{-5}$. We also ignore all other states outside of the $P$ manifold since their separation from the ground $S$ states are too far detuned from the applied laser fields to give appreciable Stark shifts.

Assuming that 200 mW of time-averaged power is focused down to a 3 $\upmu$m waist, the differential second-order Stark shift on the qubit splitting is $\delta\omega^{(2)}=E^{(2)}_{10}-E^{(2)}_{00}=-7.3$ kHz.

We will show that there is a fourth-order effect that can be much larger than the differential second-order Stark shift when using a frequency comb for specific polarizations of the beam. An intuitive understanding can be gained by considering that any two pair of comb teeth, $k_0$ and $k_1$, have a beat-note frequency $(k_0-k_1)2\pi\nu_{rep}$. If the bandwidth of the pulse is large enough, then there will be beat-notes that are close to the ground state hyperfine splitting. Assuming that none are on resonance, these off-resonant couplings can have a large effect on the ground states, as much as three orders of magnitude larger than the differential AC Stark shift.

\FigureOne

We first calculate the fourth-order Stark shift in the simplified case of just two comb teeth and one excited state of the $\ybion{171}$ level structure (see Fig. \ref{fig:AtomicLevelDiag}), equivalent to two phase coherent continuous wave (CW) beams in a three level system. Let the excited state $\ket{e}$ have frequency splitting $\omega_e$ from the $\ket{00}$ ground state, and the absolute frequencies of the comb teeth $k_0$ and $k_1$ be $\omega_0$ and $\omega_1$ respectively. Also, let the polarization of each tooth, $i$, be defined as $\hat{\epsilon}^i=\hat{\epsilon}=\epsilon_-\hat{\sigma}_-+\epsilon_0\hat{\pi}+\epsilon_+\hat{\sigma}_+$ with $|\epsilon_-|^2+|\epsilon_0|^2+|\epsilon_+|^2=1$ where $\hat{\sigma}_-,\hat{\pi}$, and $\hat{\sigma}_+$ are the polarization basis in the frame of the atom. In the rotating frame of the electro-magnetic fields of the laser, we can write the Hamiltonian 
\eq\label{eq:H0V}
\begin{split}
\mathcal{H}=&\mathcal{H}_0+V\\
=&\delta\ket{10}\bra{10}+\Delta\ket{e}\bra{e}\\
&+\frac{\Gamma^0}{2}\ket{00}\bra{e}+\frac{\Gamma^1}{2}\ket{10}\bra{e}+h.c.
\end{split}
\eeq
where $\mathcal{H}_0$ contains the diagonal terms and $V$ includes the off-diagonal terms induced by the laser, $\delta=\omega_{HF}-(\omega_0-\omega_1)$, $\Gamma^i=g_0C(\hat{\epsilon}^i)$ is the resonant Rabi frequency from beam $i$ with a dipole coupling matrix element $C(\hat{\epsilon}^i)$ for polarization $\hat{\epsilon}^i$. The fourth-order correction $E_n^{(4)}$ to the ground state energy levels, from perturbation theory, has the following form:
\eq\label{eq:E4Theory}
\begin{split}
E_n^{(4)}=&\sum_{j,l,m\neq n}\frac{V_{n,m}V_{m,l}V_{l,j}V_{j,n}}{E_{n,m}E_{n,l}E_{n,j}}-\frac{|V_{n,j}|^2}{E_{n,j}}\frac{|V_{n,m}|^2}{(E_{n,m})^2}\\
&-2 V_{n,n} \frac{V_{n,m}V_{m,l}V_{l,n}}{(E_{n,l})^2 E_{n,m}} +V_{n,n}^2\frac{|V_{n,m}|^2}{(E_{n,m})^3}.
\end{split}
\eeq
Here $j,l,m,$ and $n$ each represent different energy levels, $V_{a,b}=\bra{a}V\ket{b}$, $E_{a,b}=E^{(0)}_a-E^{(0)}_b$ is the unperturbed energy differnece between the states $\ket{a}$ and $\ket{b}$. Applying this to the Hamiltonian above, the last two terms are zero since $V$ has no diagonal terms leaving the fourth-order Stark shifts of the qubit levels,
\eq\label{eq:E4f}
\begin{split}
E_{00}^{(4)}=&-\frac{|\Omega|^2}{4\delta}\\
E_{10}^{(4)}=&\frac{|\Omega|^2}{4\delta}.
\end{split}
\eeq
In these expressions, we assume $\delta\ll\Delta$ and $\Gamma_0\sim\Gamma_1$. We also parametrize $\Omega=\Gamma_0\Gamma_1/2\Delta$, which is the resonant ($\delta=0$) stimulated Raman Rabi frequency.

\FigureTwo

The above derivation is valid for any three level system. We now include the more complete case in $\ybion{171}$ where all excited states with major contributions, namely the $^2P_{1/2}$ and $^2P_{3/2}$ manifolds, are considered. Calculating the fourth-order Stark shift on any state $\ket{n}$ reduces to computing its shift due to all other states coupled via a two-photon Raman process by fields at frequencies $\omega_0$ and $\omega_1$. In $\ybion{171}$, this means we must consider all hyperfine ground states. The two Zeeman states, $\ket{F=1,m_f=\pm 1}$, of the ground state manifold, denoted as $\{\ket{11},\ket{\var{1-1}}\}$ have a Zeeman splitting $\omega_{Zee}/2\pi\approx \pm7$ MHz under a magnetic field of approximately 5 Gauss. To caluclate the fourth-order Stark shift, we sum over all states $\ket{a}\neq\ket{n}$,
\eq\label{eq:E4bicrhom}
\begin{split}
E^{(4)}_n&=\sum_{a\neq n}\frac{\Omega_{n,a}^2}{4\delta_{n,a}}\\
\end{split}
\eeq
where $\Omega_{n,a}$ is the two-photon Rabi frequency between $\ket{n}$ and $\ket{a}$, $\delta_{n,a}=\omega_a-(\omega_0-\omega_1)$, and $\omega_a=E^{(0)}_a-E^{(0)}_n$. Computing all of the relevant Rabi frequencies $\Omega_{n,a}$ under the same assumptions as in Eq. \ref{eq:twoStark} \cite{Wineland2003}, we find
\eq\label{eq:Rabi}
\begin{split}
\Omega_{00,10}&=\left(\epsilon^{0}_-\epsilon^{1}_--\epsilon^{0}_+\epsilon^{1}_+\right)\Omega_0\\
\Omega_{00,\var{1-1}}&=-\left(\epsilon^{0}_-\epsilon^{1}_\pi+\epsilon^{0}_\pi\epsilon^{1}_+\right)\Omega_0\\
\Omega_{00,11}&=\left(\epsilon^{0}_+\epsilon^{1}_\pi+\epsilon^{0}_\pi\epsilon^{1}_-\right)\Omega_0\\
\Omega_{10,\var{1-1}}&=\left(\epsilon^{0}_-\epsilon^{1}_\pi+\epsilon^{0}_\pi\epsilon^{1}_+\right)\Omega_0\\
\Omega_{10,11}&=\left(\epsilon^{0}_+\epsilon^{1}_\pi+\epsilon^{0}_\pi\epsilon^{1}_-\right)\Omega_0.
\end{split}
\eeq
Here $\Omega_0=\frac{g_0^2}{6}\left(\frac{1}{\Delta}+\frac{1}{\omega_F-\Delta}\right)$ and $g_0^2=\gamma^2\bar{I}/2I_0$. From Eq. \ref{eq:Rabi}, we see that if $\hat{\epsilon}=\hat{\sigma}_\pm$, the Rabi frequency $\Omega_{00,10}$ is maximized and equal to $\Omega_0$. If instead $\hat{\epsilon}=\hat{\beta}\equiv 1/2\hat{\sigma}_-+1/\sqrt{2}\hat{\pi}+1/2\hat{\sigma}_+$, then $\Omega_{00,10}=0$ while all other Rabi frequencies are equal to $\Omega_0/\sqrt{2}$. These polarizations are the two which provide the largest Rabi frequencies, while all others have smaller effective Rabi rates, so we dwell on these two cases. An important note is that in the case of $\hat{\epsilon}=\hat{\beta}$, $E_{10}^{(4)}=0$ because the shifts from $\ket{11}$ and $\ket{\var{1-1}}$ are equal and cancel each other. 

We now compute the differential fourth-order Stark shift on the qubit states $\ket{10}$ and $\ket{00}$, $\delta\omega^{(4)}=E_{10}^{(4)}-E_{00}^{(4)}$,
\eq
\delta\omega^{(4)}= 
\begin{cases}
\frac{\Omega_0^2}{2\delta_{00,10}} & \mbox{when }\hat{\epsilon} =\hat{\sigma}_\pm\\
\frac{\Omega_0^2}{8}\left(\frac{1}{\delta_{00,11}}+\frac{1}{\delta_{00,\var{1-1}}}\right) & \mbox{when } \hat{\epsilon} =\hat{\beta}.\\
\end{cases}
\eeq

Finally, we generalize to incorporate all possible pairs of comb teeth. The two-photon Rabi frequency for any two comb teeth $k_0$ and $k_1$, where $k_1-k_0=l$ is $\Omega_n=g_{k_0} g_{k_{0}+l}/2\Delta\approx\Omega_0 \sech(\pi l\nu_{rep}\tau)$ \cite{Mizrahi2013}.  Let $j$ be defined such that $|\omega_a-2\pi j\nu_{rep}|$ is minimized, assuming that it is nonzero. If we now plug this into Eq. \ref{eq:E4bicrhom} summing over all comb teeth, we find
\eq\label{eq:E4Comb}
\begin{split}
E^{(4)}_n&=\sum_{a\neq n}\frac{\Omega_{n,a}^2}{4}\sum_{k=-\infty}^{\infty}\frac{\mbox{sech}^2((j+k)\pi\nu_{rep}\tau)}{\delta_{n,a}-k (2\pi\nu_{rep})}\\
&=\sum_{a\neq n}\mathcal{C}_{n,a}\frac{\Omega_{n,a}^2}{4\delta_{n,a}}
\end{split}
\eeq
where $\delta_{n,a}=\omega_a-j(2\pi\nu_{rep})$, and 
\eq\label{eq:Cna}
\mathcal{C}_{n,a}=\sum_{k=-\infty}^{\infty}\frac{\mbox{sech}^2((j+k)\pi\nu_{rep}\tau)}{1-k(2 \pi\nu_{rep})/\delta_{n,a}}.
\eeq
Because the denominator in Eq. \ref{eq:Cna} grows rapidly with $k$, only the closest few beatnotes are important, and as long as $2\pi\nu_{rep}\gg\omega_{Zee}$, then $E_{10}^{(4)}$ remains zero for $\hat{\epsilon}=\hat{\beta}$. The differential fourth-order Stark shift then becomes
\eq\label{eq:finald4}
\delta\omega^{(4)}= 
\begin{cases}
\mathcal{C}_{00,10}\frac{\Omega_0^2}{2\delta_{00,10}} & \mbox{when } \hat{\epsilon} =\hat{\sigma}_\pm\\
\frac{\Omega_0^2}{8}\left(\frac{\mathcal{C}_{00,11}}{\delta_{00,11}}+\frac{\mathcal{C}_{00,\var{1-1}}}{\delta_{00,\var{1-1}}}\right) & \mbox{when } \hat{\epsilon} =\hat{\beta}.\\
\end{cases}
\eeq
Assuming the same parameters as with the second-order Stark shift (200 mW of time-averaged power focused down to a 3$\upmu$m waist), we find that the fourth-order shift is 
\eq\label{eq:vald4}
\delta\omega^{(4)}/2\pi= 
\begin{cases}
23 \mbox{ MHz} & \mbox{when } \hat{\epsilon} =\hat{\sigma}_\pm\\
12 \mbox{ MHz} & \mbox{when } \hat{\epsilon} =\hat{\beta}.\\
\end{cases}
\eeq
This result is 1000 times larger than the differential second-order Stark shift for the same parameters. Comparing the fourth and second-order expressions, we find that $\delta\omega^{(4)}/\delta\omega^{(2)}\propto g_0^2/(\omega_{HF}\delta)$, clearly defining the regime where the fourth-order shift dominates. The second-order shift only becomes larger with a thousand-fold reduction in the laser intensity, corresponding to an applied shift below 10 Hz. Since the differential fourth-order shift can easily be made very large as shown above, it is a practical means to control a large number of qubits.

\section{Experimental Setup}

The laser used to generate the fourth-order Stark shift is a mode-locked, tripled, ND:YV$\mbox{O}_4$ \cite{Paladin2008}, at 355nm with a repetition rate of $\nu_{rep}=120$ MHz, a maximum average power of $\bar{P}=4$W, and a pulse duration of $\tau\approx 14$ps, giving a bandwidth of about 70 GHz. These parameters are well-suited for the $\ybion{171}$ system since the laser bandwidth covers the qubit splitting but does not give rise to appreciable spontaneous emission from the excited states \cite{Campbell2010}. 

The optical access of our current vacuum chamber restricts the polarization of the Stark shifting beam since the magnetic field is orthogonal to all viewports, prohibiting the use of pure $\sigma_{\pm}$ light. However, as discussed earlier, the differential fourth-order Stark shift has two possible polarizations with large shifts for a single beam: the first is pure $\sigma_{\pm}$, the second is $\hat{\epsilon}=\hat{\beta}\equiv 1/2\hat{\sigma}_-+1/\sqrt{2}\hat{\pi}+1/2\hat{\sigma}_+$. We use the elliptical polarization which slightly reduces the maximum shifts applicable, but does not require pure $\sigma_{\pm}$. 

The small spot size required to individually apply a shift to each qubit is achieved by using the imaging objective designed for qubit state readout. Since the cycling transition of $\ybion{171}$ is $369$ nm and the center wavelength of the modelocked laser is $355$ nm, we use a Semrock dichroic beam combiner (LP02-355RU-25) for separating the 355 nm laser from the resonant light at 369 nm (Fig. \ref{fig:BeamPath}). Guided by simulations of the optical system in the commercial ray-tracing software, Zemax \cite{Zemax2013}, we focus the 355 nm light down to a less than 3 $\upmu$m horizontal waist using an objective lens with a 0.23 numerical aperture.

In order to address each ion in a chain of up to 10 sites, we use an acousto-optical defelector (AOD, Brimrose CQD-225-150-355). Since the AOD is not imaged, it maps the rf drive frequency to ion position and the rf power of that drive frequency to the applied intensity. The rf control is implemented using an arbitrary waveform generator (AWG, Agilent M8109A), because it allows precise, easy, and arbitrary control while being easily reconfigurable. The differential fourth-order Stark shift is a direct change in the energy splitting of the qubit, so unlike in stimulated Raman processes, phase coherence does not require optical phase stability or even rf phase stability, but only depends on the integrated time-averaged intensity. Thus phase-coherent control only requires timing resolution better than the period of the differential fourth-order Stark shift, which is easily achieved with the AWG. The AWG also allows the application of many frequencies to the AOD, which will Stark shift multiple ions simultaneously. Additionally, the AWG gives arbitrary amplitude control of each frequency, providing time-dependent amplitude modulation of the four photon Stark shift.

\FigureThree

Due to the quadratic dependance of the differential fourth-order Stark shift on intensity, when we divide the optical power across $N$ ions, each ion's fourth-order Stark shift is diminished by a factor $N^2$,
\eq
\delta\omega^{(4)}(ion)=\mbox{max}(\delta\omega^{(4)})/N^2.
\eeq
In order to recover a linear dependance, we ``raster", or rapidly sweep, the beam position from site to site across the chain. If this rastering occurs much faster than the dynamics of the system, then the effective fourth-order shift can be safely time-averaged, yielding
\eq\label{eq:raster}
\delta\omega^{(4)}(ion)=\mbox{max}(\delta\omega^{(4)})\frac{m t_0}{T}
\eeq
where $m$ is the number of raster cycles in the total elapsed time $T$ and $t_0$ is the time the light is applied to each ion in a single cycle. In order for the raster to be fast enough to justify averaging the Stark shift, the length of each raster cycle, $Nt_0$, must be small compared to the total elapsed time $T=N m t_0$. Substituting into Eq. \ref{eq:raster},
\eq\label{eq:rasterf}
\delta\omega^{(4)}(ion)=\mbox{max}(\delta\omega^{(4)})\frac{1}{N}
\eeq
which recovers a linear dependance on the system size. In Fig. \ref{fig:RasterSeq}, we show a diagram of an example raster sequence. The limitation on this technique is how small $t_0$ can be made. In our case, $t_0$ is limited by the rise time of the AOD, which is approximately 50 ns, which is still fast compared to $N/\delta\omega^{(4)}$ and very fast when compared to a mechanical deflector rise time. 

\section{Experimental Demonstration}

\FigureFour

\FigureFive

Using Ramsey spectroscopy \cite{Ramsey1990}, we measure the total Stark shift on the qubit splitting from the applied light. A quadratic dependance on the intensity distinguishes the fourth-order Stark shift from the typical linear dependance of the second-order AC Stark shift (Eq. \ref{eq:finald4}). By measuring the total shift as a function of applied time-averaged optical power, the data in Fig. \ref{fig:shiftandwaist}a demonstrates that the observed shift is consistent with the $\bar{I}^2$ dependance of the fourth-order Stark shift. 

By translating the ion through the beam, we measure the horizontal beam waist by fitting the resulting Stark shift to the square of a Gaussian distribution (Fig. \ref{fig:shiftandwaist}b):
\eq
\delta\omega^{(4)}(\Delta x)=\delta\omega^{(4)}(0)\left(e^{-2 \Delta x^2/\sigma^2}\right)^2.
\eeq
We measure the horizontal waist to be $\sigma=2.68\pm 0.03$ $\upmu$m. This small waist allows for independent control of qubits. In Fig. \ref{fig:resolution}a, we show how qubit 5 can be driven in a ten ion system with only minimal crosstalk of approximately 2$\%$ on the adjacent spins (ions 4 and 6). In this configuration, the ions are separated by 2.76 and 2.64 $\upmu$m respectively. By increasing the distance between ions, we can decrease the crosstalk on adjacent spins. For example, in a system of two spins separated by 7 $\upmu$m, we individually drive each ion with the cross-talk $\le 2\times10^{-5}$ over a time $t=30\times 2\pi/\delta\omega^{(4)}$. 

As indicated above, the rf drive frequency maps to position at the ion chain, while the small spot size allows for individual control of the ions. In Fig. \ref{fig:resolution}b, we show this mapping in a chain of ten ions by scanning the drive frequency of the AOD while fixing the rf power and time. The difference in the applied fourth-order Stark shift of each ion is due to the rf bandwidth of the AOD, since the diffraction efficiency is lower at the extremes of the  bandwidth. In the current optical setup, a change of 10 MHz to the drive frequency corresponds to a displacement of approximately 3.4 $\mu$m along the ion chain.

This control enables the preparation of arbitrary, high-fidelity product states when the individual addressing beam is used in conjunction with global qubit operations from the Raman beams. In Fig. \ref{fig:stateprep}, we illustrate a pulse sequence used to generate a product state. This method, effectively a Ramsey sequence, is used to prepare a spatially-alternating spin state, which is the most difficult state to produce since it is the most susceptible to crosstalk. We observe a fidelity of $87\%$ for the desired state, which includes all state preparation and measurement (SPAM) errors. This fidelity is consistent with a SPAM error of only 1.4\% on each ion, where the residual infidelities come from intensity noise, the small inter-ion crosstalk of the individual addressing beam, and the ion detection error.

\FigureSix

\section{Conclusion}

The freedom and control afforded by an individually addressed, Stark-shifting beam opens many possibilities that were previously inaccessible to clock state qubits. One such new application is that we can now apply site dependent transverse magnetic fields to an interacting Ising spin system \cite{Smith2015}. Since the strength of each field is controlled by the rf amplitude from the AWG, we are able to quickly generate hundreds of different random instances of individual ion fields in a reproducible way. Furthermore, this technique enables dynamic individual control, enabling quantum simulations of interesting systems such as loops with non-zero magnetic flux \cite{Grass2015}.

The primary limitation in the current apparatus is the intensity applied to each ion, especially those on the edge of the chain due to the bandwidth of the AOD. The maximum intensity on each ion is simply
\eq
I_{ion}=\frac{2 \pi \bar{P} (\mbox{NA})^2}{\lambda^2}DE_{\nu}
\eeq
where $\bar{P}$ is the time-averaged power into the AOD, NA is the objective numerical aperture, $\lambda$ is the wavelength of the light, $DE_{\nu}$ is the diffraction efficiency of the AOD at the drive frequency, $\nu$, corresponding to the ion position. By enlarging the NA of the objective lens, the intensity applied to each ion would  greatly increase while simultaneously lowering the inter-ion crosstalk. Further, improving the diffraction efficiency and bandwidth of the AOD will allow more ions to be addressed. By implementing changes on both of these elements, we should be able to address 20+ ions without difficulty.

In this work we demonstrate that a large Stark shift can be generated on a clock state qubit with modest laser powers via a fourth-order light shift using an optical frequency comb. We show that by focusing this light, it can be used to rotate individual qubits with low crosstalk, create arbitrary product states, and generate site-specific terms in a model Hamiltonian. These new tools are important additions to the quantum toolbox and may be integral to future developments in quantum information.

\bibliography{../Bibliography/lotsofrefs}
\end{document}